\documentclass[conference]{IEEEtran}
\IEEEoverridecommandlockouts

\usepackage{url}
\usepackage{graphicx}
\usepackage{subfigure}
\usepackage{amsfonts,latexsym}
\usepackage{amsmath}
\usepackage{microtype}
\usepackage{booktabs}
\usepackage{subfigure}
\usepackage{tabularx}
\usepackage{indentfirst} 
\usepackage{multicol}
\usepackage{pdfpages}
\usepackage{multirow}
\usepackage{listings}
\usepackage{xcolor}
\usepackage{xspace}
\usepackage{adjustbox}

\newcommand{\sysname}{\textsc{IU-GUARD}\xspace}

\usepackage{amsmath}
\usepackage{lipsum} 
\usepackage{array}
\usepackage{multirow}
\usepackage{multicol}
\usepackage{algorithm}
\usepackage{algorithmic}
\usepackage{enumitem}
\usepackage{mathrsfs}
\usepackage{amsmath}

\usepackage{graphicx}
\usepackage{subcaption}
\usepackage{cite}
\usepackage{amsmath,amssymb,amsfonts}
\usepackage{algorithmic}
\usepackage{graphicx}
\usepackage{textcomp}

\begin{document}

\title{\sysname: Privacy-Preserving Spectrum Coordination for Incumbent Users under Dynamic Spectrum Sharing
}

\author{\IEEEauthorblockN{
Shaoyu Li\IEEEauthorrefmark{1},
Hexuan Yu\IEEEauthorrefmark{1},
Shanghao Shi\IEEEauthorrefmark{1},
Md Mohaimin Al Barat\IEEEauthorrefmark{1}, \\
Yang Xiao\IEEEauthorrefmark{2},
Y. Thomas Hou\IEEEauthorrefmark{1}, and
Wenjing Lou\IEEEauthorrefmark{1}
}
\IEEEauthorblockA{\IEEEauthorrefmark{1} Virginia Tech, VA, USA}
\IEEEauthorblockA{\IEEEauthorrefmark{2} University of Kentucky, KY, USA}

\thanks{Accepted at the IEEE International Conference on Communications (ICC), 2026.}

}

\maketitle

\begin{abstract}

With the growing demand for wireless spectrum, dynamic spectrum sharing (DSS) frameworks such as the Citizens Broadband Radio Service (CBRS) have emerged as practical solutions to improve utilization while protecting incumbent users (IUs) such as military radars. However, current incumbent protection mechanisms face critical limitations. The Environmental Sensing Capability (ESC) requires costly sensor deployments and remains vulnerable to interference and security risks. Alternatively, the Incumbent Informing Capability (IIC) requires IUs to disclose their identities and operational parameters to the Spectrum Coordination System (SCS), creating linkable records that compromise operational privacy and mission secrecy.

We propose \sysname, a privacy-preserving spectrum sharing framework that enables IUs to access spectrum without revealing their identities. Leveraging verifiable credentials (VCs) and zero-knowledge proofs (ZKPs), \sysname allows IUs to prove their authorization to the SCS while disclosing only essential operational parameters. This decouples IU identity from spectrum access, prevents cross-request linkage, and mitigates the risk of centralized SCS data leakage. We implement a prototype, and our evaluation shows that \sysname achieves strong privacy guarantees with practical computation and communication overhead, making it suitable for real-time DSS deployment.

\end{abstract}


\section{Introduction}

The rapid growth of mobile broadband, Internet of Things (IoT), and mission-critical wireless services has substantially increased demand for radio spectrum. To address this scarcity, U.S. regulators such as the Federal Communications Commission (FCC) and National Telecommunications and Information Administration (NTIA) introduced dynamic spectrum sharing (DSS) frameworks to enhance spectrum utilization by enabling commercial systems to opportunistically access underutilized frequency bands while ensuring strict protection for incumbent users (IUs), including military radars and satellite systems.

A representative DSS deployment is the Citizens Broadband Radio Service (CBRS, 3.55–3.70 GHz), which employs the Spectrum Access System (SAS) for spectrum coordination. 
SAS implements a three-tier access model: IUs (Tier 1) retain unconditional protection; Priority Access License (PAL) users (Tier 2) obtain licensed access in specific regions; and General Authorized Access (GAA) users (Tier 3) opportunistically utilize available channels on a non-interference basis. 
By dynamically reallocating PAL and GAA users when IUs become active, SAS ensures efficient utilization and reliable protection of federal operations. 
The NTIA also plans to open several new spectrum bands for DSS, including 3.1–3.45 GHz, 5030–5091 MHz, and 7125–8400 MHz, emphasizing the growing role of dynamic sharing in spectrum policy \cite{NTIA2023SpectrumStrategy}.

\begin{figure}[t]
    \centering
    \subfigure[ESC: sensor-based detection.]{
        \label{subfig:esc}
        \includegraphics[width=0.95\columnwidth]{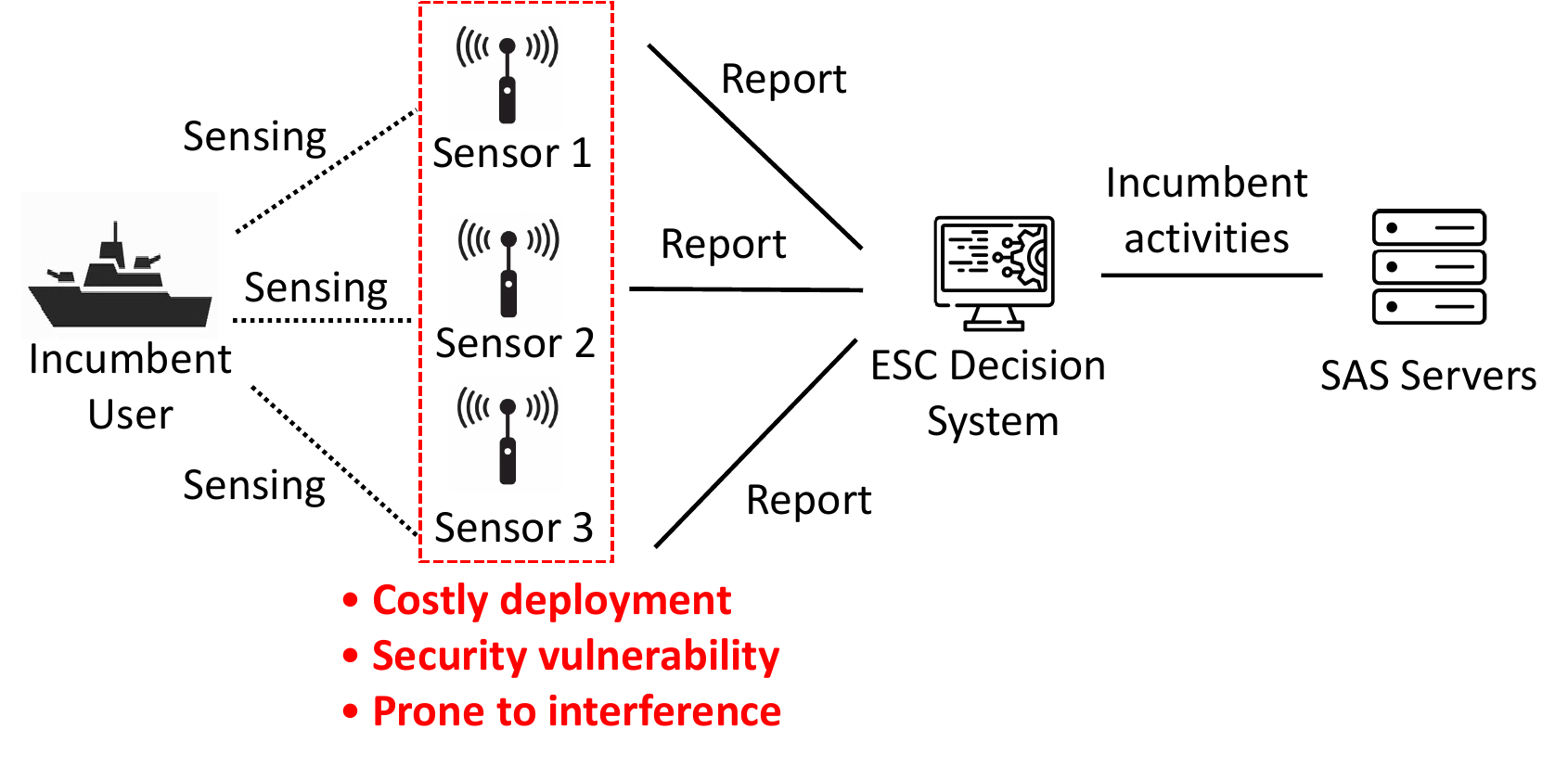}}

    \subfigure[IIC: identity-based reporting.]{
        \label{subfig:iic}
        \includegraphics[width=0.95\columnwidth]{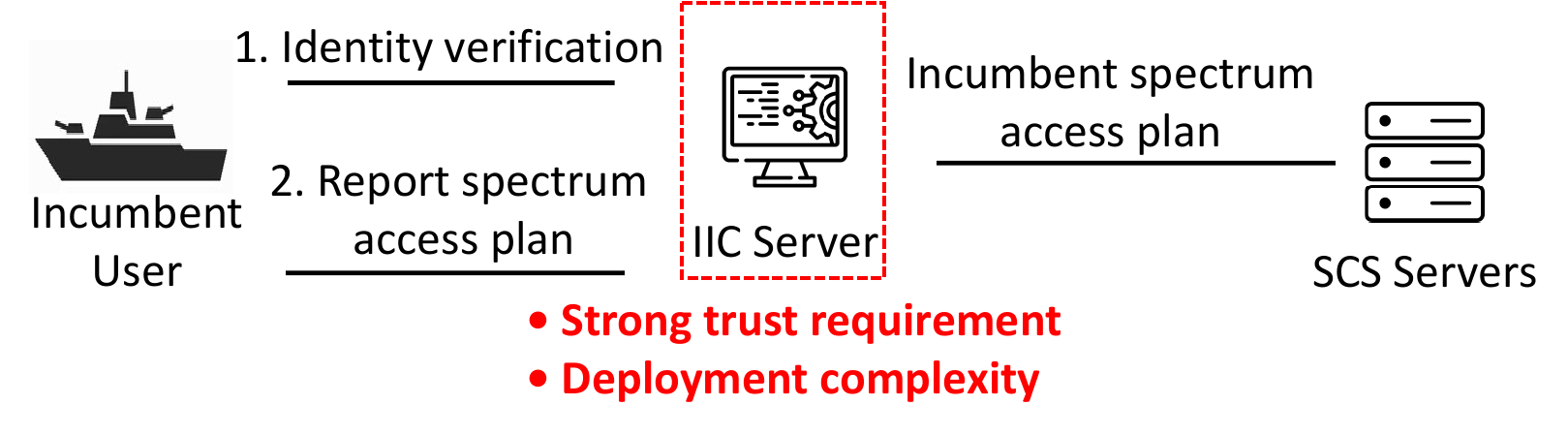}}

    \subfigure[\sysname: privacy-preserving informing.]{
        \label{subfig:iuguard}
        \includegraphics[width=0.95\columnwidth]{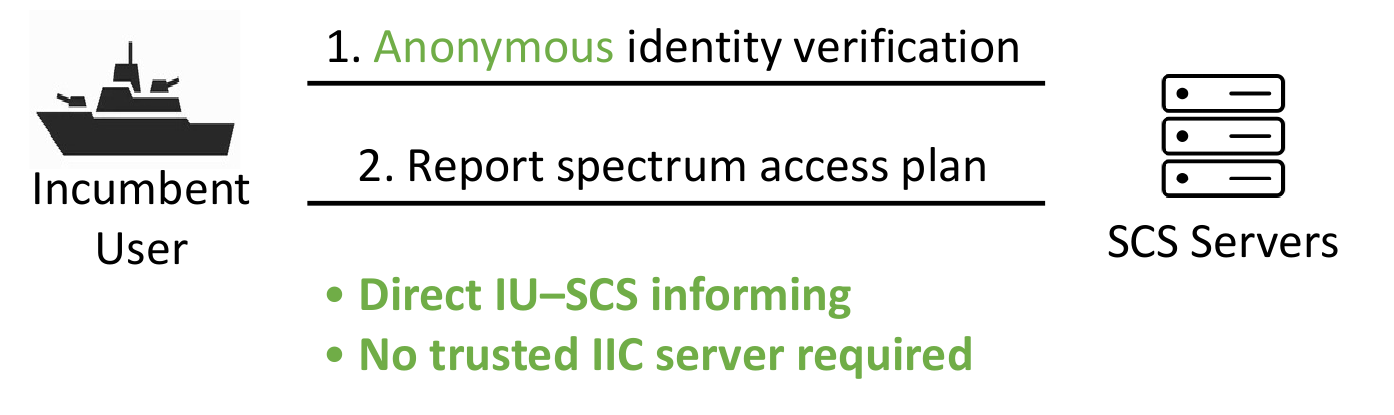}}

    \caption{Evolution of incumbent protection under DSS.}
    \label{fig:comparison}
\end{figure}

To protect IUs, two mechanisms have been standardized, as shown in Fig.~\ref{subfig:esc} and Fig.~\ref{subfig:iic}. The Environmental Sensing Capability (ESC) relies on coastal sensors to detect naval radar activity and notify the SAS of IU presence. However, ESC deployments are costly, vulnerable to attacks~\cite{shi2021challenges}, and prone to interference from neighboring bands~\cite{WINNF-RC-1016}. Alternatively, the Incumbent Informing Capability (IIC) enables IUs to directly report their activity to the Spectrum Coordination System (SCS, similar to SAS in CBRS)~\cite{ntiaIIC2021}. Although IIC provides more reliable end-to-end reporting, it introduces critical privacy risks. IUs must disclose their identities and operational parameters to the SCS, where these records are persistently stored.
However, most SCS operators are commercial entities and thus cannot be assumed trusted in handling IU data.
If compromised, the database could reveal sensitive activity traces, mission schedules, and device roles, leading to significant security risks. A privacy-preserving variant was proposed in which a centralized trusted IIC server acts as an intermediary between IUs and the SCS, verifying IU identities while concealing them from the SCS, but it has not been adopted in practice due to strong trust assumptions and limited deployability.

To address these privacy and deployability limitations, we propose \sysname, a practical privacy-preserving coordination framework that leverages verifiable credentials (VCs) and zero-knowledge proofs (ZKPs) to enhance incumbent privacy in dynamic spectrum sharing. 
Our approach requires neither disclosing identities to the SCS for verification nor relying on a centralized trusted server for authentication. 
In \sysname, IUs prove their authorization to the SCS using VCs and ZKPs without revealing their identities. 
Each IU obtains a VC from a trusted authority (e.g., FCC) and subsequently generates verifiable presentations (VPs) for each spectrum access request. 
The SCS can thus verify authorization while observing only the essential operational parameters in plaintext, effectively decoupling IU identity from spectrum access information. 
Even in the event of data leakage, \sysname ensures that no sensitive identity information can be inferred, as VPs are unlinkable and reveal no correlation across access requests. 
This design remains fully compatible with existing SCS workflows while avoiding the privacy risks of IIC.
A high-level comparison of these mechanisms is illustrated in Fig.~\ref{fig:comparison}.

\noindent\textbf{Contributions.} The main contributions are summarized:
\begin{itemize}
    \item We identify the privacy limitations of existing incumbent protection (ESC and IIC) in DSS and analyze their insufficiency in safeguarding sensitive federal operations.
    \item We propose a novel framework that leverages VCs and ZKPs to decouple incumbent identity from operational parameters. Our design allows IUs to prove their authorization without identity disclosure, while enabling SCS to allocate spectrum efficiently.
    \item We implement and evaluate the proposed framework, demonstrating that it achieves strong privacy with practical computation and communication overheads.
\end{itemize}

\section{Background and Related Work}

\subsection{Dynamic Spectrum Sharing and Incumbent Protection}

DSS improves spectrum utilization by allowing commercial users to opportunistically access federally allocated bands when temporarily unused by IUs. To ensure coexistence, the FCC and NTIA developed centralized coordination mechanisms that assign spectrum access and protect IUs from interference. For example, in the CBRS band, the SAS monitors IU activity and reassigns commercial users when IU becomes active.

To enable timely awareness of IU 
spectrum activity, ESC and IIC are standardized to notify the coordination system when IUs access a band. ESC relies on coastal sensors to detect naval radar emissions and report IU activity, but its deployment is costly, complex, and vulnerable to spoofing, jamming, and physical tampering~\cite{shi2021challenges}. Sensors near adjacent bands may also experience blocking interference from strong neighboring signals~\cite{WINNF-RC-1016}. In contrast, IIC eliminates the need for sensors by allowing IUs to directly report spectrum usage to the SCS~\cite{ntiaIIC2021}. While improving reporting accuracy and timeliness, IIC introduces critical privacy risks: IUs must disclose identifying and operational information to the SCS, where these records are persistently stored. If compromised, adversaries could infer mission schedules or device roles, causing severe information leakage and potential national security risks. A privacy-preserving variant using a trusted intermediary was proposed but has not been deployed due to limited scalability and strong trust assumptions.

\subsection{Related Work on Privacy-Preserving Spectrum Sharing}

Spectrum coordination servers (SAS in CBRS or SCS in other bands) are typically operated by commercial entities and assumed to be semihonest—they manage spectrum correctly but may infer sensitive information from user data. Protecting privacy in such settings requires balancing allocation efficiency against confidentiality, especially when operational data remains exposed during coordination.

Research on privacy-preserving spectrum sharing has largely focused on cryptographic computation. Early systems such as p2-SAS~\cite{dou2016p} employ homomorphic encryption based on the Paillier cryptosystem to execute spectrum management entirely in the encrypted domain. While effective in concealing user data from honest-but-curious servers, the high computational and communication overhead limits scalability and real-time deployment. PeDSS~\cite{li2019pedss} enhances practicality by combining homomorphic proxy re-encryption with differential privacy: incumbent reports are encrypted, while secondary users obfuscate their locations using differentially private fuzzy coordinates. This hybrid design removes the need for an online trusted party and achieves sub-millisecond latency with kilobyte-level message sizes. Pri-Share~\cite{yu2024pri} further integrates threshold-based private set union (PSU) with homomorphic encryption to enable multi-party spectrum allocation across SAS servers, ensuring FCC-compliant coordination while preserving both usage confidentiality and participant anonymity. Several works specifically target IU privacy. Lin et al.~\cite{lin2019preserving} obfuscate IU locations by adjusting antenna parameters. SZ-SAS~\cite{zabransky2019incumbent} processes encrypted operational data within the SAS to mitigate inference attacks. PriDSA~\cite{li2018preserving} and IP-SAS~\cite{dou2017preserving} employ proxy re-encryption and additive homomorphic encryption to secure safe-zone and E-Zone information against curious or colluding SAS operators.

Although these systems offer strong privacy guarantees, most depend on heavy cryptographic primitives or centralized trust, limiting their deployability in real-world DSS environments. This motivates the need for lightweight, verifiable frameworks that protect IU privacy while remaining fully compatible with existing coordination infrastructures.
\section{System Overview and Preliminaries}
\label{sec:prelim}

\subsection{System Overview}


\subsubsection{System Stakeholders}

Our system involves three key stakeholders:
\textbf{Credential Authority (CA)} is a trusted federal entity (e.g., FCC or DoD), responsible for issuing VCs to legitimate IUs. 
Before credential issuance, the CA verifies each IU’s identity and regulatory compliance based on registration records and certifies its authorization to operate within protected spectrum bands. 
Once issued, the credential enables the IU to anonymously prove its authorization to the SCS without revealing its actual identity.
\textbf{Incumbent Users (IUs)} are federal entities authorized to access spectrum resources through the SCS. 
Prior to credential issuance, each IU registers its operational profile with the DoD, including identifying information and technical parameters such as device type, antenna characteristics (e.g., gain, orientation, and height), maximum transmit power, authorized frequency range, and system type. \textbf{Spectrum Coordination System (SCS)} is a centralized coordination service responsible for managing dynamic spectrum access and enforcing protection for IUs. 

\subsubsection{Threat Model}

We assume the SCS operates under an honest-but-curious adversarial model: it correctly executes protocols in compliance with regulatory policies but may attempt to deanonymize IUs from processed data. 
This assumption is realistic, as SCS servers are often operated by third-party commercial entities that may not be fully trusted by IUs. 
In contrast, the CA is fully trusted—it issues authentic and privacy-preserving credentials to legitimate IUs and is assumed not to collude with any SCS operator or external adversary. We assume all communications are protected by secure channels.

\subsubsection{System Architecture}

\begin{figure}[t]
\centering
\includegraphics[width=0.99\linewidth]{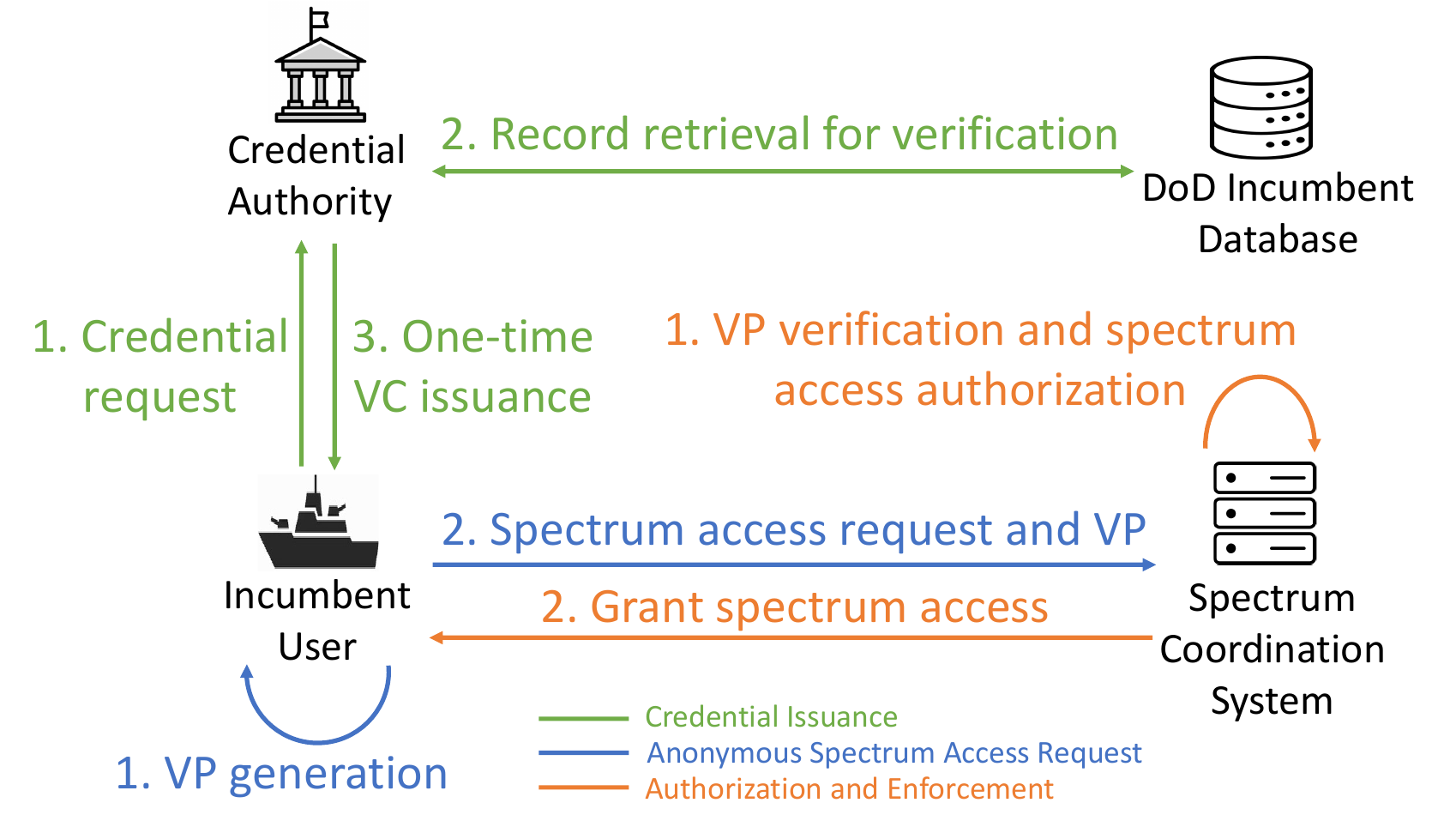}
\caption{\sysname system architecture.}
\label{fig:system_architecture}
\end{figure}

\sysname leverages VCs and ZKPs to enable privacy-preserving spectrum access authorization, as shown in Fig.~\ref{fig:system_architecture}. 
It allows IUs to prove their authorization to operate within permitted spectrum bands without disclosing their identities. 
Unlike IIC-based mechanisms that require identity disclosure or depend on a fully trusted server for verification, \sysname enables anonymous and verifiable interaction through unlinkable access requests.
The system operates in three phases:

\noindent\textbf{Phase 1: Credential Issuance.}  
Before accessing the spectrum, each IU submits a credential request to the CA. The CA verifies the IU’s identity against the DoD database and issues a VC encoding both identity and operational attributes. This credential issuance is a one-time process.

\noindent\textbf{Phase 2: Anonymous Spectrum Access Request.}  
To request spectrum access, an IU derives a VP from its VC and attaches a ZKP proving that the credential is valid and that the requested frequency bands fall within authorized limits. The IU also sends a plaintext spectrum access request containing essential metadata such as time, location, and frequency bands.

\noindent\textbf{Phase 3: Authorization and Enforcement.}  
The SCS verifies the VP using the CA’s public key but learns no identity information, as all presentations are unlinkable and privacy-preserving. Upon successful verification, the SCS grants spectrum access and reallocates lower-tier users if necessary.  

\subsubsection{Security Goals}

\noindent\textbf{Authentication and Authorization:}  
Only legitimate IUs possessing valid VCs issued by the CA can access spectrum resources.  
The SCS verifies each access request through ZKPs to confirm the IU’s authorization without direct identity disclosure, ensuring that each IU can only operate within its designated frequency bands.

\noindent\textbf{Anonymity:}  
Adversaries cannot identify the real identity or type of an IU submitting a spectrum access request.  
The anonymous VP contains no plaintext identity attributes that could reveal or infer the IU’s identity. Specifically, they cannot link the anonymous VP submitted during the access request to any record.

\noindent\textbf{Unlinkability:}  
Multiple spectrum access requests, even if submitted by the same IU at different times or locations, cannot be linked together.  
Each VP is randomized and cryptographically independent, preventing correlation or behavioral inference attacks.

\subsection{Preliminaries}

\subsubsection{Verifiable Credentials}
Verifiable Credentials (VCs)~\cite{baldimtsi2013anonymous,boneh2004short,camenisch2016anonymous} enable privacy-preserving authentication through selective disclosure. 
A VC system involves three entities: the \textit{issuer}, \textit{holder}, and \textit{verifier}. 
The issuer attests to the holder’s attributes by signing a credential using cryptographic primitives such as BBS+ signatures. 
The holder can later derive a VP by randomizing the credential’s signature and generating a ZKP to demonstrate possession of valid attributes without revealing unnecessary information.
This provides unlinkability across multiple presentations, ensuring that verifiers cannot correlate VPs issued from the same credential.
VCs have been standardized by the W3C~\cite{w3VerifiableCredentials} and IETF~\cite{ietfSDJWTbasedVerifiable}, and serve as the foundation of Decentralized Identifiers (DIDs)~\cite{W3CDecentralizedIdentifiers}, which are widely adopted in different applications~\cite{yu2024aaka,li2026fc}.
\subsubsection{Zero Knowledge Proofs}
Zero Knowledge Proofs (ZKPs)~\cite{goldreich1994definitions} enable a prover to demonstrate possession of certain
knowledge (e.g., a valid credential) without revealing the underlying secret. 
When combined with VCs, ZKPs support selective disclosure, allowing a holder to prove authorization while keeping identity information hidden and credentials unlinkable. 
A specialized form of ZKPs, known as \textit{range proofs}~\cite{bunz2018bulletproofs}, further allows verification that a numerical attribute lies within an authorized interval without exposing its actual value. 
\sysname leverages these proofs to enforce operational compliance, e.g., verifying that an IU’s transmission frequency remains within its assigned band. The ZKP mechanism implemented in \sysname is based on non-interactive zero-knowledge (NIZK) protocols, specifically utilizing BBS+ signature-based proofs.

\section{System Design}
\label{sec:system_design}

\begin{figure}[t]
\centering
    \subfigure[Credential issuance.]{
        \label{subfig:Credential_issue}
        \includegraphics[width=0.90\columnwidth]{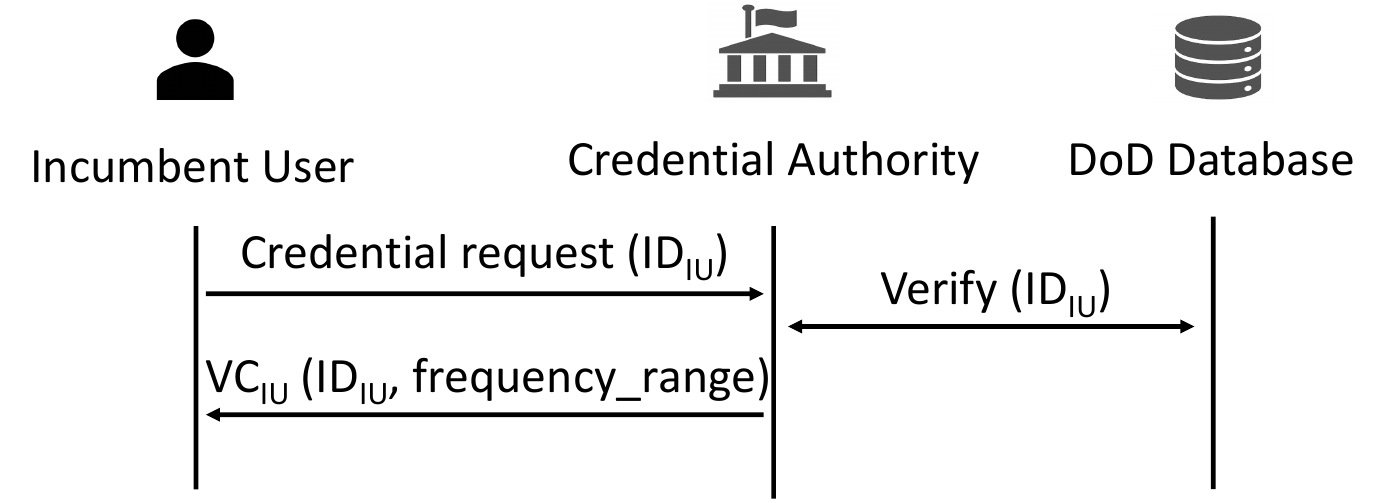}
    }

    \subfigure[Spectrum access.]{
        \label{subfig:Credential_Verify}
        \includegraphics[width=0.90\columnwidth]{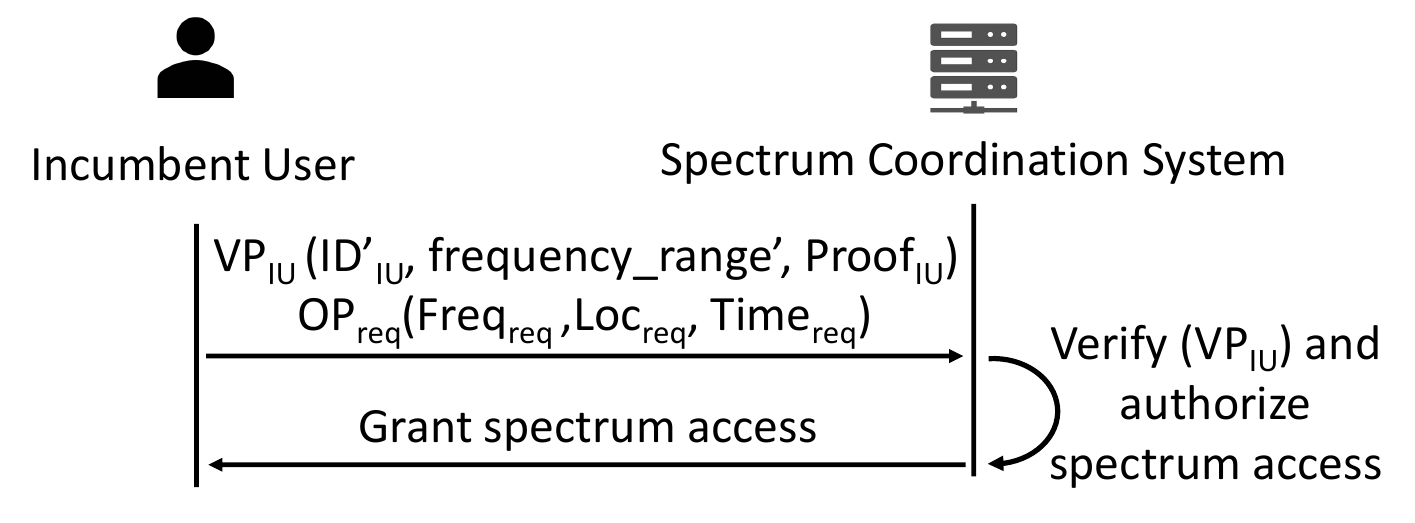}
    }

    \caption{Credential issuance and spectrum access.}
    \label{fig:Credential_protocol}
\end{figure}

The protocol workflow of \sysname consists of three main phases: credential issuance, anonymous spectrum access request, and authorization enforcement. Fig.~\ref{subfig:Credential_issue} depicts the credential issuance process, while Fig.~\ref{subfig:Credential_Verify} illustrates both the access request and authorization phases.

\subsection{Credential Issuance}
Each DoD-operated system (e.g., radar, satellite terminal, or base station) is pre-registered in the DoD spectrum management database, which records operational attributes such as device type, antenna configuration, authorized frequency range, and operating region.
When an IU submits a credential request, the CA retrieves the corresponding record, verifies its authenticity, and issues a VC that encodes certified attributes:
\begin{equation}
VC_{IU} = (ID_{IU}, freq\_range)
\end{equation} where
\begin{equation}
freq\_range = [f_{\text{low}}, f_{\text{high}}]
\end{equation}
Here, \(ID_{IU}\) denotes the IU’s pseudonymous identifier, and \(freq\_range\) specifies its authorized band.
The credential is securely returned to the IU and stored locally.
This one-time, offline issuance process involves no interaction with the SCS, ensuring that sensitive identity data remains completely isolated from coordination entities.
This credential then serves as the foundation for subsequent anonymous access verification.

\subsection{Anonymous Spectrum Access Request}
\label{phase2}

This phase enables anonymous interaction between IUs and the SCS.
After obtaining a valid credential, an IU can anonymously request spectrum access from the SCS.
Unlike IIC-based mechanisms that require identity disclosure for verification, \sysname allows an IU to prove its authorization without revealing any identifying information.
The IU derives a VP from its credential by randomizing the signature and embedding a ZKP demonstrating that the credential was legitimately issued by the CA:
\begin{equation}
VP_{IU} = (ID_{IU}', freq\_range', Proof_{IU})
\end{equation}

Here, attributes with a prime symbol (e.g., \(ID_{IU}'\), \(freq\_range'\)) are cryptographically hidden from the SCS, and \(Proof_{IU}\) proves that the IU holds a valid credential without revealing its true identity.

In addition, the IU includes minimal operational metadata in plaintext for coordination:
\begin{equation}
OP_{req} = (Freq_{req}, Loc_{req}, Time_{req})
\end{equation} where
\begin{equation}
Freq_{req} = [f_{\text{low}}^{req}, f_{\text{high}}^{req}]
\end{equation}

The proof \(Proof_{IU}\) further attests that the requested frequency range lies within the IU’s authorized band:
\begin{equation}
Proof_{IU}\left(Freq_{req} \subseteq freq\_range\right)
\end{equation} where
\begin{equation}
(f_{\text{low}} \leq f_{\text{low}}^{req}) \land (f_{\text{high}}^{req} \leq f_{\text{high}})
\end{equation}

Since all VPs are randomized, no two access requests can be linked to the same IU, ensuring both anonymity and unlinkability across sessions.

\subsection{Authorization and Enforcement}
\label{phase3}
Upon receiving \(VP_{IU}\) and \(OP_{req}\), the SCS verifies \(Proof_{IU}\) using the CA’s public key.
During this process, the SCS learns only the operational parameters disclosed in plaintext but no identity information about the IU.
If verification succeeds, the SCS grants temporary access to the requested frequency range \(Freq_{req}\) within the IU’s authorized allocation.
To maintain incumbent protection, the SCS may preempt or reassign lower-tier commercial users occupying overlapping channels.
Only minimal metadata necessary for coordination (e.g., granted frequency, time, and location) are retained, and no linkable identifiers are stored.
Even if the SCS database is later compromised, an adversary cannot trace or correlate individual IU operations.
Through this phase, \sysname achieves secure and privacy-preserving incumbent access:
IUs are verifiably authenticated while their operational anonymity and unlinkability remain fully protected.
This allows the SCS to fulfill regulatory protection requirements without compromising user privacy or system deployability.

\section{Security Analysis}
\label{sec:security_analysis}

\noindent\textbf{Anonymity.}
\sysname enables IUs to access the spectrum without revealing their real-world identities to the SCS. 
Each IU generates a VP by randomizing its credential signature and attaching a ZKP of validity. 
The VP conceals all sensitive attributes while proving compliance with authorization policies. 
Because every VP is unlinkable to the underlying credential, the SCS cannot directly deduce the IU’s identity even after observing multiple access requests. 
No pre-registration or static identifier is shared with the SCS, and only minimal operational metadata (frequency, location, time) is revealed in plaintext for coordination purposes. 
For mobile incumbents, such as aircraft or tactical systems, the dynamic nature of access requests further reduces the risk of re-identification by SCS.

\noindent\textbf{Unlinkability.}
Each spectrum access request employs a VP with independent cryptographic randomness and commitments. 
Even when derived from the same credential, two VPs are statistically indistinguishable and cannot be linked to each other or to the original credential. 
This prevents both passive eavesdroppers and honest-but-curious SCS operators from correlating access events across time or geographic regions, effectively mitigating behavior inference and tracking attacks.

\noindent\textbf{Authenticity and Authorization.}
Only IUs possessing valid credentials issued by the CA can generate a verifiable presentation that passes SCS verification. 
The embedded ZKP attests that the VC was issued by a legitimate CA and the requested operational parameters (e.g., frequency range) lie within authorized bounds. 
This guarantees that spectrum access is restricted to certified incumbents while preventing unauthorized entities from obtaining protected-band privileges.

\noindent\textbf{Resistance to Impersonation.}
Each VC is cryptographically bound to a link secret known only to the legitimate IU, preventing reuse or forgery by adversaries. 
Even if an attacker compromises an old VP or intercepts protocol messages, the randomness of each presentation and ZKP ensures that new proofs cannot be reconstructed or cloned. 
Thus, impersonation or credential replay attacks are
effectively prevented under our threat model.

\noindent\textbf{Protection from SCS Inference Attacks.}
Under the honest-but-curious model, the SCS faithfully enforces spectrum coordination but may attempt to infer additional information from received data. 
In \sysname, all sensitive attributes are cryptographically hidden behind ZKPs, and no static identifiers are disclosed during protocol execution. 
The SCS learns only the minimum information required for access decisions, ensuring compliance with functional requirements while preserving incumbent privacy. 
This design provides a realistic and deployable defense against privacy leakage from third-party-operated SCS infrastructures.

\section{Evaluation}

\subsection{Implementation Setup}
\label{sec:implementation}

We implement a prototype of \sysname\ comprising three subsystems: a CA server, an IU client, and an SCS server. The CA issues and signs VCs, the IU generates VPs and ZKPs for spectrum access requests, and the SCS verifies these proofs and enforces authorization.
For comparison, we develop a baseline DSS model that follows the standard IIC workflow, where the IU authenticates to the IIC server using plaintext credentials, and the IIC forwards access requests to the SCS without cryptographic protection.

The implementation of VCs leverages the Hyperledger \texttt{AnonCreds} library~\cite{anoncredsrs}, which conforms to the W3C Verifiable Credentials standard~\cite{w3VerifiableCredentials}. 
A local Hyperledger Indy pool is employed as the public ledger for credential schema registration.
We deploy four servers on Amazon Web Services (AWS), representing the CA, the SCS, the spectrum database, and the baseline IIC server. 
Each server runs on an \texttt{AWS c6i.3xlarge} instance with Ubuntu~22.04 LTS. 
The IU client is implemented on a local workstation with 128\,GB RAM and an Intel Xeon E5-2620\,v4 CPU (32 cores, 2.10\,GHz), running Ubuntu~22.04.4 LTS. 
All communications between the IU, CA, SCS, the database, and the baseline IIC are protected using TLS~1.3 connections, ensuring confidentiality and integrity across all channels. 
Both the communication workflow and cryptographic operations are implemented in Python.

The CA server maintains a verifiable credential schema on the Indy ledger and issues credentials to registered IUs. Each IU client derives \(VP_{IU}\) from its locally stored \(VC_{IU}\), randomizes the credential signature, and generates a proof \(Proof_{IU}\) demonstrating that its requested frequency band lies within the authorized range. 
The SCS server verifies \(VP_{IU}\) and \(Proof_{IU}\) using the CA’s public key, and authorizes access. In the baseline design, the IIC authenticates the IU using account–password methods. Upon successful authentication, the IIC forwards only the IU’s operational parameters to the SCS while concealing the user’s identity information.

\begin{figure*}[t]
    \centering
    \subfigure[Computation time breakdown.]{
        \label{Computation time breakdown of IU-GUARD}
        \includegraphics[width=0.3175\textwidth]{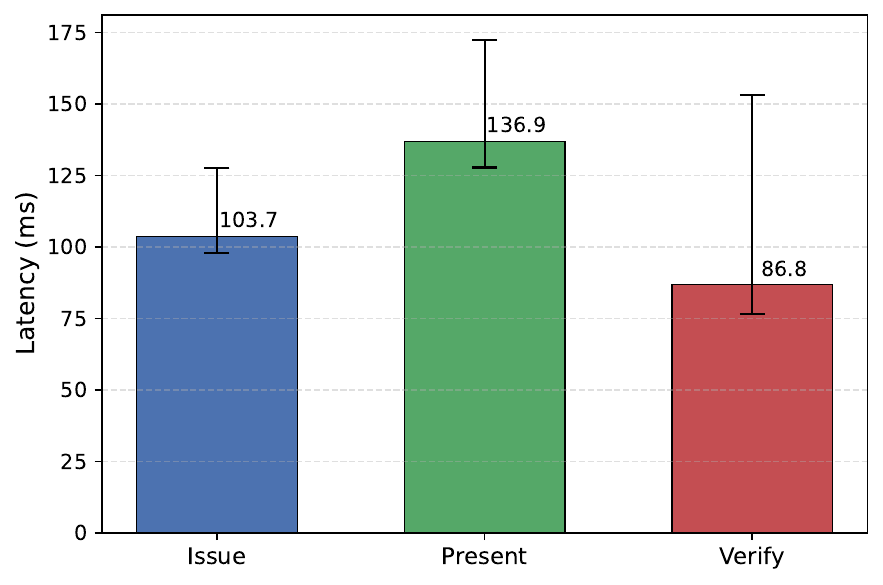}}
    \subfigure[Latency under varying numbers of users.]{
        \label{subfig:latency}
        \includegraphics[width=0.296\textwidth]{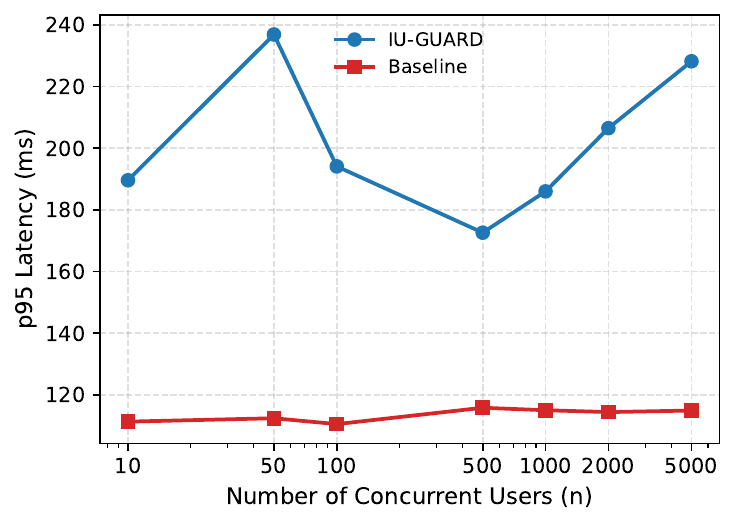}}
    \subfigure[Throughput under varying numbers of users.]{
        \label{subfig:throughput}
        \includegraphics[width=0.30\textwidth]{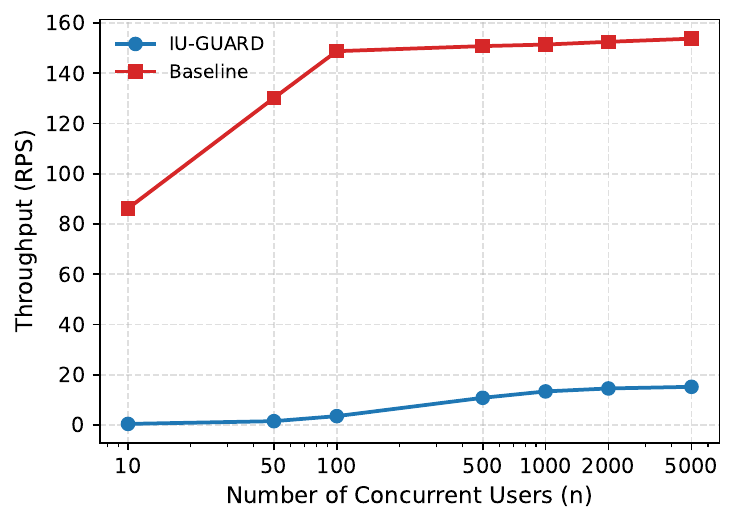}}
    \caption{Performance evaluation of IU-GUARD, including computation time and system scalability in latency and throughput.}
    \label{fig:evaluation}
\end{figure*}

\subsection{Evaluation}
\noindent\textbf{Credential and Presentation Overhead.}
We evaluate the data size overhead introduced by credentials and presentations in \sysname. The issued VC averages 5.33 KB, while each VP, including ZKP parameters and randomized commitments, averages 23.47 KB. These message sizes represent a modest increase compared to plaintext authentication in the baseline but remain within acceptable limits, imposing no noticeable impact on transmission.

\noindent\textbf{Computation Performance.}
We evaluate the computational latency of the three VC phases—Issue, Present, and Verify—over 100 trials, capturing the mean, median (p50), and tail (p95) latencies. As shown in Fig.~\ref{Computation time breakdown of IU-GUARD}, credential issuance introduces a moderate one-time cost, while presentation generation and verification complete within 175~ms at the 95th percentile. These results demonstrate the feasibility of integrating VC/ZKP primitives for real-time DSS authorization.

\noindent\textbf{Communication Performance.}
TABLE~\ref{tab:communication_performance} summarizes the communication latency comparison between the baseline and \sysname.
We evaluate three key processes: credential issuance, identity verification, and authorization.
During credential issuance, the CA retrieves IU records, constructs credentials, and applies BBS+ signatures. \sysname\ requires 128.4~ms on average, compared to 49.7~ms in the baseline without cryptographic signing. Since issuance occurs only once per IU, this one-time cost has negligible impact on real-time operations.
For identity verification, the IU submits a VP with a ZKP, which then verified by the SCS. In the baseline, plaintext credentials are transmitted for verification. \sysname\ achieves a total latency of 105.9~ms, compared to 15.2~ms in the baseline, representing a modest but acceptable increase for real-time coordination.
Finally, in the authorization stage, we measure the end-to-end latency from spectrum request submission to access grant. 
The baseline first verifies the IU’s identity within the IIC, then transmits the authentication result to the SCS, which performs direct access assignment. \sysname\ incorporates ZKP verification and policy enforcement before granting access. The total authorization latency in \sysname\ is 278.4~ms, compared to 130.3~ms in the baseline, both remaining well within practical DSS timing requirements.

\begin{table}[ht]
\centering
\caption{Latency comparison between baseline and \sysname.}
\label{tab:communication_performance}
\renewcommand{\arraystretch}{1.19}
\begin{tabular}{l|c|c}
\hline
 & \textbf{Baseline} & \textbf{\sysname} \\
\hline
Credential issuance & 49.7\,ms & 128.4\,ms \\
Identity verification & 15.2\,ms & 105.9\,ms \\
Authorization & 130.3\,ms & 278.4\,ms \\ 
\hline
\end{tabular}
\end{table}

\noindent\textbf{Scalability.}
We evaluated the scalability of \sysname\ under increasing numbers of concurrent IU access requests, compared with the baseline. In \sysname, the SCS processes concurrent spectrum access requests and VPs, while in the baseline, the IIC server performs plaintext authentication and forwards access requests to the SCS. The number of concurrent users is set to 10, 50, 100, 500, 1000, 2000, and 5000. Fig.~\ref{subfig:latency} and Fig.~\ref{subfig:throughput} show the p95 latency and throughput results. Overall, \sysname\ achieves stable latency and good scalability with increasing concurrency, whereas the baseline attains higher throughput due to plaintext processing but without privacy protection.

In summary, these results demonstrate that IU-GUARD achieves practical latency and scalability, making it suitable for real-time DSS operations.


\section{Conclusion}
\label{sec:conclusion}

In this paper, we presented \sysname, a privacy-preserving spectrum coordination framework that enables incumbent users to directly access shared spectrum without disclosing their identities. 
By integrating Verifiable Credentials (VCs) and Zero-Knowledge Proofs (ZKPs), \sysname decouples user identity from operational attributes, allowing the Spectrum Coordination System to verify authorization while preserving anonymity and unlinkability. 
We implemented a prototype and demonstrated that \sysname achieves strong privacy guarantees with minimal computational and communication overhead, maintaining compatibility with existing SCS workflows. 
Future work will explore large-scale deployment across multi-administrator DSS environments.
\section*{ACKNOWLEDGMENT}
This work was supported in part by the Office of Naval Research under grants N00014-24-1-2730, and the National Science Foundation under grants 2433904, 2433905, 2331936, 2332675, 2312447, 2247560, 2247561, 2154929, and the Virginia Commonwealth Cyber Initiative (CCI).

\bibliographystyle{ieeetr}
\bibliography{ref}

\end{document}